\documentclass[11pt,a4paper]{article}
\usepackage{epsfig,amsmath,amssymb,array,dcolumn,subfigure,rotating}
\def\brho{{\mbox{\boldmath $\rho $}}}

\def\bnabla{{\mbox{\boldmath $\nabla $}}}
\def\ul#1#2{\textstyle{\frac#1#2}}
\title{Membrane pinning on a disordered substrate}

\author{ R. Podgornik $^{1,2}$  and P.L. Hansen $^{1,3}$ \\
$^{1}$ LPSB/NICHD, Bld. 12A  Rm. 2041, \\
                                National Institutes of Health, Bethesda, MD 20892-5626 \\
$^{2}$ Department of Physics, Faculty of Mathematics and Physics, \\
University of Ljubljana,  SI-1000 Ljubljana and  Department of Theoretical Physics, \\
				J.Stefan Institute, SI-1000 Ljubljana, Slovenia \\
$^{3}$ MEMPHYS - Center for Biomembrane Physics, Department  of Chemistry \\
University of Southern Denmark, Campusvej 55, DK-5230 Odense M, Denmark}

\begin{document}



\maketitle

\begin{abstract}
We investigate interactions between an elastic membrane and a substrate
characterized by quenched positional disorder in the height function. 
We show that the positional disorder transforms the standard secondary
DLVO minimum into two separate states: the hovering state
characterized by a planar membrane at a finite separation from the
interface and a pinned state where the membrane follows closely the
asperities of the substrate and is as a consequence quite corrugated. 
The transition between the two states is continuous and depends on the
parameters of the underlying DLVO potential as well as the parameters
describing the quenched height-height correlation function of the
substrate.

\noindent
pacs{87.16.Dg}{Membranes, bilayers, and vesicles}
\noindent
pacs{68.15.+e}{Liquid thin films}

\end{abstract}

In the DLVO theory one usually assumes that when a membrane or in
general an elastic manifold interacts with a rigid substrate, the
membrane is modelled as flat and the substrate is
envisioned as featureless \cite{wennerstrom}.  This leads in a
straightforward manner to the secondary minimum of the DLVO theory and
to an equilibrium spacing between the membrane and the substrate. 
Taking into account the elastic degrees of freedom of a membrane can
alter this picture drastically, leading to the emergence of Helfrich
interaction which may induce an unbinding transition of the membrane.  
This transition has been clearly shown to be a consequence
of the interplay between elastic fluctuations of the membrane and DLVO
interactions between the membrane and the substrate \cite{lipowsky}. 
In the present work we are motivated by recent experiments on
deposited lipid multilayers on atomically smooth vs.  rough surfaces
\cite{stephanie}.

These experiments reveal differences in equilibrium
lamellar spacings of lipid multilayers in proximity of a substrate
that seem to correlate with molecular roughness of the substrate
\cite{stephanie}.  In order to lay ground for understanding effects of
this type it is desirable to relax the model constraint of a
featureless substrate and treat it supposedly more realistically as
exhibiting (quenched) disorder in the height function.  We will
analyze the consequences of this new model of a substrate interacting
with a membrane and show that it modifies the simplified DLVO
conclusions in the sense that the original secondary DLVO minimum now
splits into two separate states, characterised by the way the membrane
is (de)coupled to the substrate.  We call these states the {\sl
hovering state} and the {\sl pinned state}.  The former one is
characterized by a membrane in a flat configuration hovering a certain
finite distance above the substrate and is directly related to the
standard DLVO secondary minimum.  The latter one is characterized by a
corrugated membrane that basically follows the asperities of the
quenched positional disorder of the substrate and is altogether
missing from the standard DLVO theory.  We stress that our description
is unifying in the sense that it combines aspects of {\sl mean-field}
theory \cite{Andelman}, which focuses on the nature of the pinned
state; and a straight-forward generalization \cite{PL} of Li and
Kardar's \cite{LiKardar} {\sl Gaussian fluctuation theory}, which allows for a hovering
state determined by the parameters of the DLVO interaction potential. 
Specifically, we find that for sufficiently strong disorder, the
membrane indeed prefers the mean-field pinned state considered by
Swain and Andelman. Because of the increased role of substrate
disorder fluctuations, in weakly disordered systems the properly self
averaged free energy can give way to a hovering state where the
membrane is depinned, residing in a minimum related to but different from the usual
DLVO minimum, as predicted by Gaussian fluctuation theory.

We introduce our approach by first specifying the model of the
membrane and the substrate that interact {\sl via} a DLVO potential
with (for matters of convenience) hydration and van der Waals terms. 
Formally our analysis owes a lot to investigations of effects of the
disorder on polymerized membranes that have been analyzed in a variety
of contexts \cite{radzihovsky}.  It is based on the assumption of a
Gaussian substrate height distribution function and the application of
the $1/d$ expansion method \footnote{Strictly speaking what we use is actually 
a $1/(d-2)$ expansion.} to evaluate the complicated partition
function stemming from the replicated Hamiltonian.

The Hamiltonian of a flexible membrane described in the usual Monge
parameterization $u(\brho)$, where $\brho = (x,y)$ is a 2D coordinate
vector, above a substrate characterized by the height function
$\zeta(\brho)$, where $\zeta(\brho)$ is a quenched disorder field
describing the profile of the substrate, is composed of three
contributions.  First of all we have the elastic energy of the
membrane, then the interaction free energy per unit surface area
between the membrane and the substrate $V(u(\brho) - \zeta(\brho))$ of
the general DLVO form, and finally the free energy contribution of an
external force per unit surface area $\pi$ pushing the membrane
towards the substrate.  The total free energy or equivalently the
mesoscopic Hamiltonian thus assumes the form
\begin{equation}
	{\cal H}[u(\brho)] = \ul12\int\!\!\!\int\!\!  d^{2}\brho
	d^{2}\brho'~{\cal K}(\brho,\brho')u(\brho)u(\brho') + \int\!\! 
	d^{2}\brho~V(u(\brho)-\zeta(\brho)) + \int\!\!  d^{2}\brho
	~\pi(u(\brho)-\zeta(\brho)). \nonumber
\end{equation}
Usually one takes for the elastic part the classical Canham - Helfrich - Evans
{\sl ansatz} ${\cal K}(\brho, \brho') = \bnabla^{4} K_{c}
\delta^{2}(\brho - \brho')$, where $K_{c}$ is the bending modulus of
the membrane, while the interaction part is composed of the attractive
and repulsive DLVO ingredients.  Since the interaction part of the
Hamiltonian is in general non-linear it is convenient to introduce the
following new variable $ B(\brho) = (u(\brho)- \zeta(\brho))^{2}$ at
every $\brho$ {\sl via} a functional constraint \cite{Hansen}
\begin{eqnarray}
	\label{cons}
	\delta{\cal H}[B(\brho), g(\brho)] &=& \ul12\int\!\! 
	d^{2}\brho ~ g(\brho)\left( (u(\brho)- \zeta(\brho))^{2} -
	B(\brho)\right). \nonumber
\end{eqnarray}
Clearly the auxiliary field $g(\brho)$ plays a role akin to the
self-energy part of the Green function.  The two auxiliary fields
$B(\brho)$ and $g(\brho)$ just introduced, play the role of fixing the
local constraint $(u(\brho)- \zeta(\brho))^{2} = B(\brho)$.  They also
show up in the partition function where one has to eventually take the
trace over these auxiliary fields together with $u(\brho)$.  The
partition function can thus be written as
\begin{eqnarray*}
     {\cal Z}[\zeta(\brho)] &=& \int {\cal D}u(\brho) {\cal D}B(\brho)
     {\cal D}g(\brho) ~e^{-\beta ({\cal H}[u(\brho)] + \delta{\cal H}[B(\brho),
     g(\brho)])}. \nonumber
\end{eqnarray*}
The average over the quenched disorder distribution, assuming the
self-averaging property of the free energy, is defined to be of the form
\begin{equation}
	\overline{(\dots)} = \int {\cal D}\zeta (\brho) (\dots)
	{\cal P}[\zeta(\brho)], \nonumber 
\end{equation}
where by assumption the disorder probability distribution function
${\cal P}(\zeta(\brho))$ that characterises the quenched disorder in
the height function of the substrate is given by a Gaussian {\sl
ansatz}
\begin{equation}
	{\cal P}[\zeta(\brho)] =
	\exp{\left(-\ul12 \int\!\!\!\int\!\!  d^{2}\brho
	d^{2}\brho' ~{\cal G}(\brho,
	\brho')\zeta(\brho)\zeta(\brho')\right)}.  \nonumber
\end{equation}
The free energy, after being averaged also over the quenched disorder
distribution, is obtained finally as
\begin{equation}
	{\cal F} = - kT ~\overline{\log{{\cal
	Z}[\zeta(\brho)]}} = - kT \lim_{n\longrightarrow 0}
	\frac{\overline{{\cal Z}^{n}[\zeta(\brho)]} - 1}{n}.
	\label{freeen}
\end{equation}
With these preliminaries the free energy Eq.  \ref{freeen} can be
evaluated {\sl via} the standard Edwards-Anderson replica trick
\cite{Dotsenko}, where the replicated Hamiltonian ${\cal H}_{n}$ is
composed of the replicated elastic term
\begin{equation}
	\ul12 \sum_{i=0}^{n}
	\int\!\!\!\int\!\!  d^{2}\brho d^{2}\brho' ~{\cal K}(\brho,
	\brho')u_{i}(\brho) u_{i}(\brho'), \nonumber
\end{equation}
where $i$ is the index of the replica, the replicated constraint on the
variable $B_{i}(\brho)$ that now reads
\begin{equation}
	\ul12 \sum_{i=0}^{n} \int\!\!  d^{2}\brho ~ g_{i}(\brho)\left(
	(u_{i}(\brho)- \zeta(\brho))^{2} - B_{i}(\brho)\right), \nonumber
\end{equation}
and finally of the replicated interaction and external ``source'' terms
\begin{equation}
	\sum_{i=0}^{n} \int\!\!  d^{2}\brho~ V(B_{i}(\brho)) +
	\sum_{i=0}^{n} \int\!\!  d^{2}\brho ~\pi ( u_{i}(\brho)-
	\zeta(\brho)).\nonumber
\end{equation}
Since we now have Gaussian integrals over the variables $u_{i}(\brho),
\zeta(\brho)$ we can evaluate them explicitely, while the functional
integrals over the auxiliary fields $B_{i}(\brho), g_{i}(\brho)$ can
be evaluated on the saddle-point level, with the {\sl proviso} that
there is no replica symmetry breaking.  This constitutes the essence
of the $1/d$ expansion.

After performing all the indicated integrations and taking the $n
\longrightarrow 0$ limit, the free energy can be obtained as a sum of
the mean-field part \footnote{Not to be confused with the mean-field 
approach introduced in \cite{Andelman}. The term mean-field is used 
here strictly as it pertains to the $1/d$ expansion.} and a 
fluctuation part. The mean-field part is
\begin{eqnarray*}
	{\cal F}_{0} &=& \int\!\!  d^{2}\brho ~\pi u_{0}(\brho) +
	\ul12 \int\!\!\!\int\!\!  d^{2}\brho d^{2}\brho' \chi (\brho,
	\brho') u_{0}(\brho) u_{0}(\brho') \nonumber
\end{eqnarray*}
where the mean-field $u_{0}(\brho)$ is obtained via minimization of
${\cal F}_{0}$ with $$\chi (\brho, \brho') = {\cal K}(\brho, \brho') +
\delta^{2}(\brho - \brho') g(\brho).$$ The fluctuation part of the
free energy is concurrently obtained as
\begin{eqnarray}
	{\cal F} &=& \frac{kT}{2} \int\!\!  d^{2}\brho ~{\cal
	G}^{-1}(\brho, \brho) g(\brho) - \frac{kT}{2}
	\int\!\!\!\int\!\!  d^{2}\brho d^{2}\brho' \chi^{-1}(\brho,
	\brho') {\cal G}^{-1}(\brho', \brho) g(\brho)g(\brho') - \nonumber\\
	&-& \ul12 \int\!\!  d^{2}\brho ~g(\brho) B(\brho) + \int\!\! 
	d^{2}\brho V(B(\brho)).
	\label{frend}
\end{eqnarray}
On the $1/d$ expansion level the auxiliary fields contribute only at
the saddle point.  The saddle point equations are now obtained simply
by minimizing Eq.  \ref{frend} with respect to $B(\brho)$ and $
g(\brho)$.  We will not reproduce the general rather awkward form of
these equations but will concentrate on a rather particular solution
characterized by $u_{0} = const.$ and $B = const.$, implying also $g =
const.$.  In addition to this we will limit ourselves to the
conceptually most interesting case of vanishing external confining
force, {\sl i.e.} $\pi = 0$.

Assuming that the system is homogeneous in the $\brho$ plane we can
introduce the Fourier transforms of all the relevant quantities that
allow us to write the mean-field equations in a rather simple form
\begin{equation}
	u_{0} \chi({\bf Q} = 0) = u_{0} g = 0.
	\label{MF}
\end{equation}
The saddle point for $B$ is obtained straightforwardly as
\begin{equation}
	\frac{\partial V(B)}{\partial B} = \ul12 g, \nonumber
\end{equation}
while the saddle point for $g$ can be derived in the form
\begin{equation}
	B = u_{0}^{2} + kT \sum_{\bf Q} \frac{{\cal K({\bf
	Q})}^{2}}{{\cal G}({\bf Q}) ({\cal K}({\bf Q}) + g)^{2}}.
	\nonumber
\end{equation}
This set of equations has two fundamentally different solutions
describing the state of the elastic membrane interacting with a
disordered substrate.
\begin{enumerate}
	\item a {\sl hovering state}, with $u_{0} \neq 0, g = 0$, characterized
	by
	\begin{equation}
		\frac{\partial V(B)}{\partial B} = 0 \qquad {\rm with} \qquad
		B = u_{0}^{2} + kT \sum_{\bf Q} {\cal G}^{-1}({\bf
		Q}).
	\end{equation}
	\item and a {\sl pinned state}, with $u_{0} = 0, g \neq 0$,
	characterized by
	\begin{equation}
		\frac{\partial V(B)}{\partial B} = \ul12 g \qquad {\rm with} \qquad
		B = kT \sum_{\bf Q} \frac{{\cal K({\bf Q})}^{2}}{{\cal
		G}({\bf Q}) ({\cal K}({\bf Q}) + g)^{2}}.
	\end{equation}
\end{enumerate}
Obviously, in order to progress we have to assume a certain form for
the quenched disorder correlation function.  As a simplest
approximation we take a spatially short range coupling ${\cal
G}(\brho, \brho') = G \delta^{2}(\brho - \brho')$, implying ${\cal
G}({\bf Q}) = G$.  Thus we have in the hovering state
\begin{equation}
	\frac{\partial V(B)}{\partial B} = 0 \qquad {\rm with} \qquad
	u_{0}^{2}= B - B_{c},
	\label{hover}
\end{equation}
where we introduced $B_{c} = kT \sum_{\bf Q} {\cal G}^{-1}({\bf Q}) =
\frac{kT}{G} \frac{Q_{max}^{2}}{4\pi}$, that obviously depends on the
upper wavevector cutoff in the Fourier space.  The first of the above
equations determines $B$ as a function of the parameters of the DLVO
potential ({\sl e.g}.  Hamaker constant, hydration interaction
strength etc.).  The second one gives the dependence of $u_{0}$ on
these parameters.  

In order to understand the physical nature of the two phases, we evaluate
the average of the separation between the substrate and the membrane
that can be obtained as
\begin{equation}
	S^{-1} \int\!\!  d^{2}\brho~\overline{\mathopen<
	\left(u(\brho) - \zeta(\brho) \right)\mathclose>} = u_{0}.
	\nonumber
\end{equation}
Clearly $u_{0}$ quantifies the disorder averaged separation between
the membrane and the substrate.  For a finite $u_{0}$ the membrane
hovers a finite separation away from the substrate.  In order to
characterize the hovering state further we evaluate the average square
of the deviation from a unit normal to the membrane in the $z$
direction given to the lowest order as $\delta{\bf n}(\brho) \approx
\bnabla_{\perp} \cdot u(\brho)$,
\begin{equation}
	p^{2} = S^{-1}\!\!\int\!\!  d^{2}\brho~\overline{\mathopen<
	\delta{\bf n}^{2}(\brho)\mathclose>}  = \frac{kT~g^{2}}{2 G}
	\sum_{\bf Q}\frac{Q^{2}}{({\cal K}({\bf Q}) + g)^{2}} =
	\frac{kT}{16 \pi} \frac{g}{G~K_{c}}.
	\label{psquare}
\end{equation}
Thus in the hovering state with $g = 0$ not only is the membrane
decoupled from the substrate and hovers above it, but is also flat on
the average since the mean disorder averaged squared deviation from
the normal, $p^{2}$, equals zero at a finite separation $u_{0}$ (see
figure).  
\begin{figure}[t]
\begin{center}
\epsfig{file=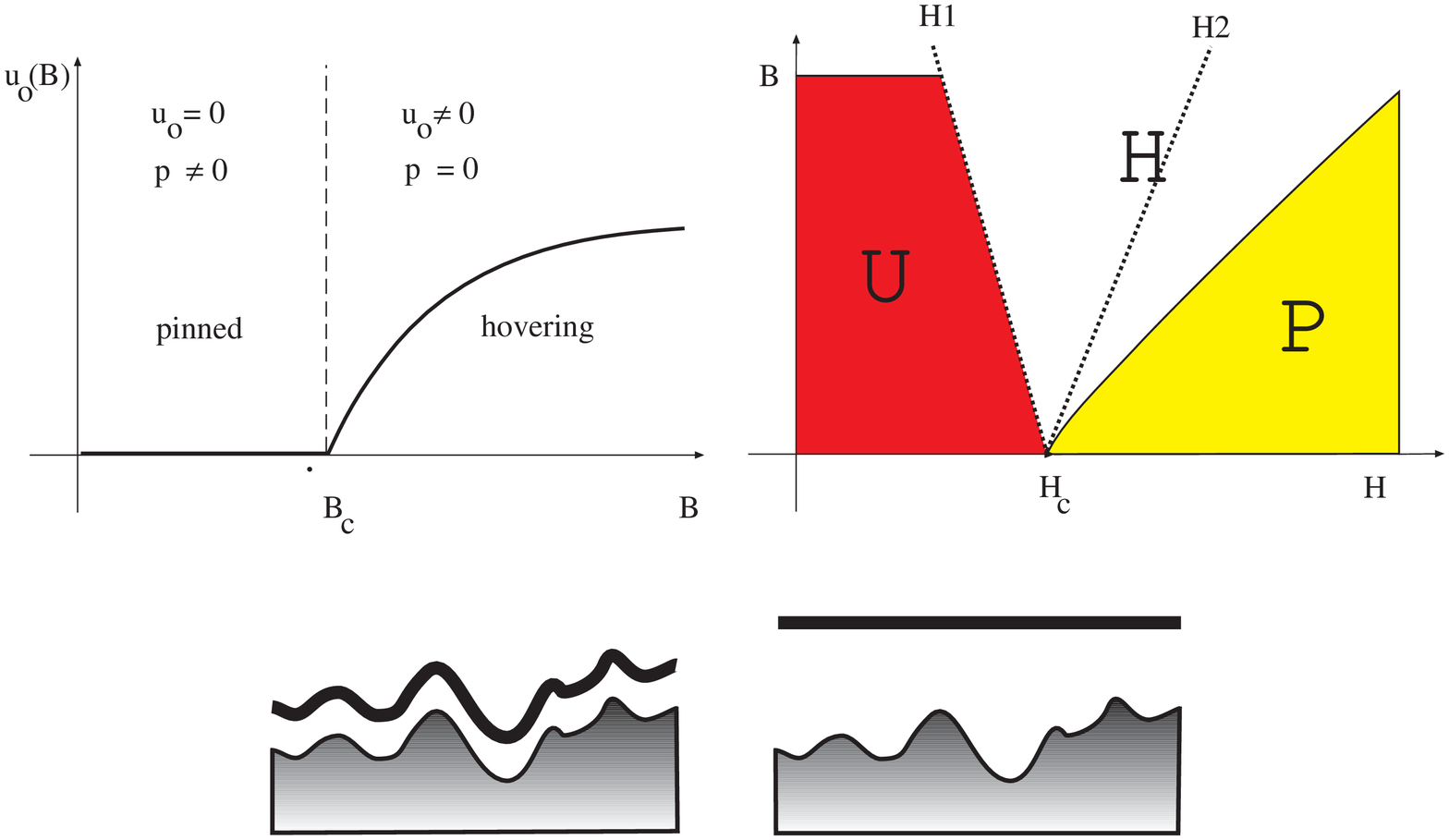, width=14cm, angle=0}
\end{center}
%
\label{schwartz}
\caption{Upper left: Schematic dependence of $u_{0}$ on the auxiliary field $B$. 
The hovering state exists only for $B > B_{c}$.  The critical value
$B_{c}$ depends on the disorder characteristics of the substrate and
the microscopic wavevector cutoff.  The hovering - pinned transition
is obviously second order in $B$, which thus plays the role of the
temperature.  Upper right: Schematic phase diagram of a membrane as a
function of the Hamaker coefficient; $V(u)$ is assumed to be
equal to the sum of hydration, van der Waals and Helfrich terms Eq. 
\ref{dis1}.  The substrate disorder introduces a hovering (H) state 
window  into the phase diagram, located between the pinned (P) and 
the unbound (U) phases. Two possible (H-U) boundaries are indicated 
by the H1 and H2 lines. The size of this window depends on the 
strength of the Hamaker coefficient and the value 
of $G$, characterizing the disorder effects.  In the case of no 
disorder ($B \longrightarrow 0$) the critical value of the Hamaker coefficient 
$H_{c}$ marks the unbinding transition of the membrane.  Bottom: A schematic
representation of the hovering (H) and the pinned (P) state of the
membrane in proximity of a disordered substrate.  The hovering state
is characterized by a slowly varying average separation
between the membrane and the substrate that does not follow closely
the local corrugations of the substrate.  In the pinned state however,
the membrane follows closely the asperities of the substrate.}
\end{figure}
Obviously the hovering solution exists only for $B > B_{c}$.  The
hovering line in the ``phase diagram'' thus ends at the value of the
interaction parameters where the solution of $\frac{\partial
V(B)}{\partial B} = 0$ also satisfies $B = B_{c}$.  $u_{0}$ thus
behaves as an order parameter of a second order phase transition, and
$B$ behaves as the temperature.

Furthermore the surface density of the adhesion free energy ($U$) in
the hovering state is given by
\begin{equation}
	U = min \left\{ \frac{\cal F}{S} \right\} = V(B) = V(u_{0}^{2} + B_{c}).
	\nonumber
\end{equation}
This obviously differs from the standard secondary minimum of the DLVO
theory, determined from $\frac{\partial V(u_{0})}{\partial u_{0}} =
0$, with $U = V(u_{0})$.  The free energy in the hovering state is
thus not given by the value of the DLVO potential at the minimum
$V(u_{0})$, but at a (much smaller) value of $V(u_{0}^{2} + B_{c})$,
since $V$ is a decreasing function of its argument above its minimum.

In the hovering state the membrane is obviously in close proximity
(``contact'') of the substrate only for a fraction of its total
surface area.  This effective contact area ($S_{c}$) can be estimated 
\cite{Andelman} from
\begin{equation}
    \frac{S_{c}}{S} = \frac{V(u_{0}^{2} + B_{c})}{V(u_{0}^{2})}.
    \nonumber
\end{equation}    
In view of the discussion presented above, this ratio is smaller then one if $V$ is a
decreasing function of its argument above its minimum value.

For the pinned state, with ${\cal K}({\bf Q}) = Q^{4} K_{c}$, we
obtain the following expression to the lowest order in the wavevector
cutoff $Q_{max}$
\begin{equation}
	\frac{\partial V(B)}{\partial B} = \ul12 g \qquad {\rm with} \qquad
	B = B_{c} - \frac{3 kT}{16 G} \sqrt{\frac{g}{K_{c}}},
	\nonumber
\end{equation}
where $B_{c}$ is defined in the same way as in the hovering state. 
The adhesion free energy in the pinned state, defined in complete
analogy with the hovering state, is given by
\begin{equation}
	U = V(B) + \frac{kT~K_{c}}{4\pi G} \left(
	\frac{g}{K_{c}}\right)^{\frac{3}{2}}
	= V(B) + \frac{16^{3}}{4\pi 3^{3}}
	\frac{K_{c} G^{2}}{(kT)^{2}} \left(  B_{c} - B\right)^{3}.
	\nonumber
\end{equation}
In the pinned state the membrane is thus coupled to the substrate and
follows it closely, being always in its close proximity since $u_{0} =
0$ (see figure).  It thus exhibits a very corrugated configuration. 
This follows again from Eq.  \ref{psquare} since in the pinned state
$p^{2} \neq 0$.  We note that for non-zero external driving force
$\pi$ the system is always in the pinned state.  

The adhesion energy in both states is nowhere in general equal
to its DLVO counterpart.  Both, the hovering as well as the pinned
states carry in the free energy the signature of the substrate
disorder in the height-height correlations.  Only in the limit of
vanishing disorder, or in the language of our model as $G
\longrightarrow \infty$, does the hovering state approach the DLVO
secondary minimum in a continuous fashion, while the pinned state
simply disappears.

In order to gain further insight into the nature of the hovering and
the pinned states we investigate the phase diagram for a particular
typical choice of the DLVO interaction potential augmented by the
Helfrich undulation interaction.  For the sake of simplicity we assume
that the DLVO part is given by the sum of the hydration and van der
Waals interactions while the Helfrich interaction is assumed to have
the same form as between a flexible membane and a flat substrate
\cite{wennerstrom,lipowsky}, thus
\begin{equation}
    V(u) = A e^{-u/\xi} - \frac{H}{12 \pi u^{2}} + \frac{6   \pi^{2}
    (kT)^{2}}{256 K_{c} u^{2}} 
    \label{dis1}
\end{equation}    
where $A$ is the magnitude and $\xi$ the range of the hydration
interaction, and treat the Hamaker constant $H$ as a variable tuning
parameter.  The above form of the total interaction energy between a
membrane and a substrate would be strictly valid only for a membrane
fluctuating near a flat substrate.  However Swain and Andelman note
\cite{Andelman} that the form of the Helfrich interaction should not
be far off from the one in the above expression even for a rough
substrate.  We assume this is the case when we analyze the phase
diagram for a membrane near a disordered substrate.  This assumption
would however have to be tested {\sl via} a more sophisticated and
hopefully more accurate approach.

Without the disorder the interaction potential Eq.  \ref{dis1} leads
to an unbinding transtion at the critical value of the Hamaker
constant equal to $H_{c} = \frac{72 \pi^{3} (kT)^{2}}{256 K_{c}}$. 
When $H$ approaches this value the secondary DLVO minimum is displaced
towards infinity and we have a continuous unbinding of the membrane. 
Adding disorder to this scenario we instead obtain a modified phase
diagram as presented on the figure.  There is now a window
corresponding to the hovering state (the former DLVO secondary
minimum) in between the pinned and the unbound state of the membrane. 
The dimensions of this window depend on the value of $\cal G$ that
characterizes the intensity of the substrate disorder.  In the limit
of no disorder $\cal G \longrightarrow \infty$ with $B_{c}
\longrightarrow 0$ the hovering window is expanded to the whole axis
$H > H_{c}$ and is tranformed back into the standard DLVO secondary
mimimum.

The disorder usually does not figure in the theories of membrane
substrate interactions.  The present work is basically a plea for a
change of this perspective.  There obviously exist phenomena, where
ignoring the disordered nature of the substrate does not, even
qualitatively, lead to the correct physical picture.  In this sense
the DLVO theory has to be ammended.

{\bf Acknowledgement}

MEMPHYS - Center for Biomembrane Physics of the University of Southern Denmark is funded by the Danish Natural Science Research Council.

\end{document}